\documentclass[traditabstract]{aa} 
\usepackage{longtable}
\usepackage{epsfig}
\usepackage{graphicx}
\usepackage{hyperref}
\usepackage{xcolor}
\usepackage{txfonts}
\usepackage{amssymb}
\usepackage{natbib}       
\usepackage{url} 
\usepackage{orcidlink}
\usepackage{hyperref}

\bibpunct{(}{)}{;}{a}{}{,}
\newcommand{\be}{\begin{equation}}
\newcommand{\ee}{\end{equation}}
\newcommand{\bea}{\begin{eqnarray}}
\newcommand{\eea}{\end{eqnarray}}

\begin{document}

\title{Neutron star atmospheres composed of fusion ashes}
\author{Valery~F.~Suleimanov\inst{\ref{in:Tub}}\orcidlink{0000-0003-3733-7267}
\and
Juri Poutanen\inst{\ref{in:UTU}}\orcidlink{0000-0002-0983-0049}
\and
Klaus Werner\inst{\ref{in:Tub}}\orcidlink{0000-0002-6428-2276}
}

\institute{Institut f\"ur Astronomie und Astrophysik, Kepler Center for Astro and Particle Physics, Universit\"at T\"ubingen, Sand 1, 72076 T\"ubingen, Germany\\ \email{suleimanov@astro.uni-tuebingen.de}
\label{in:Tub} 
\and 
Department of Physics and Astronomy, 20014 University of Turku,  Finland 
\label{in:UTU} \\
}

\date{Received 16 January 2026 / Accepted 29 April 2026}

\authorrunning{Suleimanov et al.}
\titlerunning{Neutron star atmospheres composed of fusion ashes}

\abstract
{Here we present models of hot neutron star (NS)  atmospheres consisting of thermonuclear ashes of various chemical compositions. 
These models are essential for studying thermonuclear flashes in X-ray bursting NSs in which nuclear-burning ashes are transported to the stellar surface. 
We consider four different mixtures, each dominated by helium, chromium, iron, or nickel.
In addition to the opacity sources previously used in NS atmosphere modeling, we include photoionization from excited ionic states as well as approximately 5000 spectral lines. 
We also develop a method that enables the simultaneous treatment of Compton scattering and a large number of spectral lines.
A key feature of the modeled NS atmospheres is the presence of a layer in the transition region between the optically thin and optically thick parts of the atmosphere where the radiation-pressure force increases significantly. 
This enhanced force sets an upper limit on the maximum attainable bolometric flux for a given surface gravity and chemical composition.
The emergent spectra from the computed atmospheres display pronounced absorption edges, whose energies are determined by the dominant chemical species. 
We fit the model spectra using a diluted blackbody modified by a single absorption edge, and we investigate how the fit parameters depend on both the relative bolometric flux and the chemical composition of the atmosphere.
Finally, we discuss constraints on these models imposed by the properties of X-ray bursts that exhibit absorption edges in their spectra, as observed in the systems  HETE~J1900.1$-$2455 and GRS~1747$-$312. } 


\maketitle
\nolinenumbers

\section{Introduction}

On the surfaces of many neutron stars (NSs) in low-mass X-ray binaries (LMXBs), freshly accreted material undergoes 
thermonuclear flashes. 
These systems are known as X-ray bursters \citep[see reviews by][]{1993SSRv...62..223L,2021ASSL..461..209G}. 
In some cases, the bursts become so energetic that their luminosities reach the Eddington limit. 
Because of this property, X-ray bursts provide valuable opportunities for determining NS masses and radii \citep[see, e.g.,][]{2016ARA&A..54..401O, Suleimanov.etal:16, 2018ASSL..457..185D}.

The observed X-ray spectra of thermonuclear flashes are, in almost all cases, well fitted by a blackbody \citep[see, e.g.,][]{Galloway08}. 
In reality, however, the spectra deviate from a true blackbody. In a hot NS atmosphere -- an appropriate model for the outer layers during an X-ray burst -- the spectral shape is determined primarily by Compton scattering. 
The intense exchange of energy and momentum between electrons and photons drives the radiation field toward thermodynamic equilibrium and, to first approximation, yields a spectrum resembling a diluted blackbody \citep{London86, Lapidusetal:86, 1987PASJ...39..287E}.
Specifically, the emergent flux from a NS atmosphere with effective temperature $T_{\rm eff}$ can be approximated via Planck function as 
\begin{equation}
F_E \approx w\,\pi\, B_E (f_{\rm c} T_{\rm eff}) =    w\,\pi\, \frac{2}{c^2h^2}\  \frac{E^3}{\exp(E/k_B f_{\rm c} T_{\rm eff})-1} . 
\end{equation}
 The  spectral dilution factor $w <1$ and the color correction $f_{\rm c} >1$ depend on the surface gravity parameter $\log g$, the relative luminosity $\ell = F/F_{\rm Edd}$, and the chemical composition of the atmosphere \citep{SPW11, SPW12, Nattila.etal:15}. 
Here $F_{\rm Edd} = cg/0.2(1+X)$ is the Eddington flux (i.e., the flux at which radiation pressure for Thomson scattering opacity balances the gravity),  $c$ is the speed of light, and $X$ is the hydrogen mass fraction. 
A direct consequence is that the  NS radius inferred from a simple blackbody fit $R_{\rm BB}$, underestimates the true stellar radius  $R$ also known as a local circumferential NS radius  (further just NS radius). 
For a uniformly emitting surface and negligible contamination from other X-ray-emitting components, the apparent radius $R_\infty$ is larger than the circumferential radius $R$ due to the light bending and scales as
\begin{equation}
 R_\infty = R\, (1+z) \approx w^{-1/2} R_{\rm BB},  
\end{equation} 
where $1+z = (1-2GM/c^2R)^{-1/2}$ is a correction for the gravitational redshift. 
Here $M$ is the NS gravitational mass.   

During the cooling phase of X-ray bursts occurring in the persistent hard spectral state -- when no optically thick accretion disk is present close to the NS and accretion occurs through a hot, optically thin accretion flow  \citep[see more detail in][]{SPRW11, Suleimanov.etal:16, Kajava.etal:14} -- the measured blackbody radius, $R_{\rm BB}$, evolves with decreasing luminosity in a manner consistent with the predicted variation of the spectral dilution factor $w$ from hot NS atmosphere models \citep[see, e.g.,][]{Suleimanov.etal:16}. 
This correspondence has enabled robust estimates of NS radii, which have been found to lie in the range of 11--13 km for several LMXBs \citep{Nattila.etal:16, Nattila.etal:17, Sul.etal:17, Suleimanov.etal:17}. 
Independent radius measurements derived from modeling the X-ray pulse profiles of millisecond pulsars observed with NICER have confirmed these values \citep{2019ApJ...887L..24M, 2019ApJ...887L..21R, 2021ApJ...918L..28M, 2021ApJ...918L..27R}.

Some X-ray bursts are so powerful and long-lasting that a significant fraction of the NS envelope can be expelled by a super-Eddington wind, exposing layers enriched with thermonuclear burning products \citep[see, e.g.,][]{2006ApJ...639.1018W,2010A&A...520A..81I,2017MNRAS.464L...6K,2018ApJ...866...53L}.
It is therefore interesting that NICER has detected several spectral features during powerful photospheric radius expansion (PRE) bursts from the ultracompact LMXB 4U\,1820$-$30 that may be associated with heavy elements produced in thermonuclear reactions \citep[see][and references therein]{2025ApJ...986...16J}.
The absorption and emission features were detected during the PRE phase when the photospheric radius reached 75–100~km for the adopted distance of 8.4~kpc \citep{2019ApJ...878L..27S}.
NSs in LMXBs typically rotate rapidly, with spin frequencies exceeding 200–300~Hz \citep[see, e.g.,][]{SB06,Watts2012,2017ApJ...850..106P,DiSalvo2024}.
Consequently, when the photospheric radius is close to the stellar surface, spectral features are expected to be strongly broadened by Doppler effects.
At much larger photospheric radii, however, the rotational velocity should decrease due to angular momentum conservation, making it possible for relatively narrow spectral features to appear in the X-ray spectra.

Models of super-Eddington winds from X-ray bursting NSs have been developed by many authors \citep[see, e.g.,][]{1986ApJ...302..519P,1994ApJ...433..276N,2018ApJ...863...53Y,2020A&A...638A.107H,2023A&A...678A.156H,2021ApJ...914...49G}.
Our subsequent work will be based on the results presented by \citet{2018ApJ...863...53Y}, who computed the chemical element abundances in the atmospheres of X-ray bursting NSs after the completion of the wind-driven erosion of the upper atmospheric layers.
More detailed calculations of thermonuclear reactions during X-ray bursts exist and extend to the production of heavier elements \citep[see, e.g.,][]{2004ApJS..151...75W,2008ApJS..174..261F,2010ApJS..189..204J,1960ApJ...132..883G,2023A&A...678A.156H}. 
However, most of these studies did not consider powerful super-Eddington bursts, which are required to bring thermonuclear ashes to the surface.
A second reason for adopting the ash compositions from \citet{2018ApJ...863...53Y} is the lack of reliable atomic data for elements heavier than nickel. 
Atmospheres enriched with heavier elements should undoubtedly be investigated in future studies.

It is clear that the emergent spectra of the model atmospheres enriched with heavy elements are different from the model spectra of atmospheres with solar H/He composition. 
The first attempt to model atmospheres including fusion ashes was made by \citet{Nattila.etal:15}. 
Model atmospheres with many times (up to 40) increased heavy elements abundances in comparison with solar abundances with keeping a solar H/He ratio were computed. 
These models were supplemented by atmospheric models consisting of pure iron. Each model spectrum was fitted by a diluted blackbody spectrum, and theoretical dependencies $F/F_{\rm Edd} - f_{\rm c}$ were presented for all computed spectral sequences. 

 A common feature of the obtained dependencies is a decrease in $f_{\rm c}$ with an increase 
in heavy element abundances. \citet{2017MNRAS.464L...6K} found that during the cooling phase  of one of the X-ray bursts of HETE J1900.1$-$2455, 
there was a jump in the value of the blackbody radius $R_{\rm BB}$. This jump can be explained by the fact that matter with solar H/He ratio began 
to accrete onto the surface of the NS enriched with heavy elements. The second example is an unusually long burst of GRS~1747$-$312 with a super-Eddington wind phase \citep{2018ApJ...866...53L}. 
It was only possible to describe the spectral evolution of this burst during the cooling phase by assuming that the NS atmosphere consists of pure iron.

Thus, we are confident that the surface of a NS after a powerful and prolonged thermonuclear flash can be enriched with  combustion products, and such X-ray bursts are observed. 
This potentially opens the way for the study of thermonuclear ashes on the NS surface. 
The first step towards reaching this aim is the modeling of the NS atmospheres with chemical compositions predicted by model computations of X-ray flashes.
We chose four different compositions presented by \citet{2018ApJ...863...53Y}. 
We also computed such model atmospheres with an admixture of plasma with solar He/H abundance plus reduced metal abundances in various proportions. 

In the following Sect.\,\ref{sect:method} we describe our method to construct the model atmospheres. 
In Sect.\,\ref{sect:uncertainties} we present the properties and spectral evolution of the models enriched by heavy elements. 
The results are discussed in Sect.\,\ref{sect:discussion} together with constraints on the models by two observed X-ray bursts. 
We conclude in Sect.\,\ref{sect:conclusions}.

\section{Method}
\label{sect:method}

Our approach to modeling atmospheres of hot NSs including Compton scattering is described in \citet{SPW12}. 
It is based on solving the integro-differential equation of radiation transport. 
The Compton effect is taken into account using a fully relativistic angle-dependent redistribution function \citep[see, e.g.,][]{1996ApJ...470..249P}. 
The corresponding code was written on the basis of Kurucz's  ATLAS code \citep{1970SAOSR.309.....K,1993KurCD..13.....K}.
It is based on a temperature-correction method that we modified to account for the Compton effect. 
The calculations included the 15 most abundant chemical elements, H, He, C, N, O, Ne, Na, Mg, Al, Si, S, Ar, Ca, Fe, and Ni. 
We assumed local thermodynamic equilibrium (LTE) to calculate the ionization states and number densities of excited ion levels. 
We note, however, that in previously calculated model atmospheres  \citep{SPW12, Suleimanov.etal:17} the contribution of heavy elements to the opacity was considered in a simplified manner. 
In particular, only photoionization from the ground states of the ions was taken into account \citep[see][]{1996ApJ...465..487V, 2003ARep...47..186I}, and the contribution of spectral lines was ignored.
These same simplifications were used to calculate models of atmospheres enriched with heavy elements \citep{Nattila.etal:15}. 

When computing the model atmospheres  presented here, we improved the calculation of the heavy element contributions to the opacity.
First of all, we add two elements that were not taken into consideration before, namely, Cr and Ti.  
The most detailed consideration is given to hydrogen-like and helium-like ions of heavy elements, similar to what was done when modeling NS carbon model atmospheres \citep{2014ApJS..210...13S}. 
Five excited levels of hydrogen-like ions and ten excited levels of helium-like ions were taken into account.  
Cross sections of photoionization from these levels were calculated in the approximation of a pure Coulomb potential \citep{1961ApJS....6..167K}. 
To calculate the partition functions of the ions, taking into account pressure ionization using the occupation probability formalism \citep{1988ApJ...331..794H}, 96 levels of hydrogen-like ions and 11 levels of helium-like ions are considered. 
Occupation probabilities are computed using the approach suggested by \citet{1994A&A...282..151H}. 
For the remaining ions, the statistical  weight of the ground state was considered to be the partition function.

The photoionization cross sections from excited levels of other ions are represented in the form of analytical approximations obtained by fitting the numerical results presented by the Opacity Project \citep[OP;][]{1994MNRAS.266..805S}, see \cite{2024A&A...688A..39S} for a detailed description.  
The exceptions are Cr and Ti ions with three or more electrons. 
For these elements, there are no data on photoionization cross sections in the OP databases. 
Therefore, photoionization from the excited levels was not accounted for these ions. 
Note that the approximations used to describe the opacities of lithium-like and other less charged ions do not significantly affect the results, since, under the parameters of the model atmospheres  considered, almost all chemical elements are essentially fully ionized.
  
Another important difference from previously published models is the consideration of opacity in spectral lines. 
Data on excitation energies of ion energy levels, wavelengths of transitions between levels, and transition oscillator strengths ($gf$ values or Einstein spontaneous-emission probabilities) for most ions were taken from the  CHIANTI database \citep{1997A&AS..125..149D, 2021ApJ...909...38D}.  
For a number of spectral lines (in particular, for the lines of Cr and Ti ions), data were taken from the NIST database.\footnote{https://physics.nist.gov} 
Only spectral lines in the 0.1--12~keV spectral band, which contains the main radiation flux in the considered model atmospheres, were taken into account. 
Doppler broadening, radiative damping, and Stark broadening were considered factors that broaden the spectral lines. 
For most lines, the Stark broadening was calculated using the approximation suggested by \citet{1971Obs....91..139C}.  
The Stark widths of hydrogen-like ions were calculated using the approximation proposed by \citet{1960ApJ...132..883G,1967ApJ...148..547G}. 
For this purpose, Kurucz's modified subroutine \citep{1970SAOSR.309.....K} was used. 
Spectral lines were considered in the LTE approximation. 
This means that scattering in the lines was not taken into account because the source function in that case is given by the Planck function. 
The total number of spectral lines taken into consideration is about 5000.

Usually, the opacity sampling method is employed to account for the opacity of the spectral lines. 
In this approach, we do not describe the accurate profile for every spectral line in the emergent spectrum. 
This is not possible for a significant number of lines. 
Instead, we solve the radiation transfer equation at a large number ($\sim$\,40\,000) of logarithmically equally spaced frequency points.  
However, this approach cannot be directly applied if it is necessary to also take Compton scattering into account.  
The radiation transport equation including the Compton effect  is solved in 380 spectral bands. 
To account for opacity in spectral lines, each spectral band in the 1--12~keV range is divided into 50 intervals, where the true absorption opacity (without electron scattering) is calculated. 
Then, the obtained values are averaged within a given spectral band $j$ as
\begin{equation} \label{eq:av}
       \frac{1}{\kappa_j} = \frac{1}{\Delta E_j} \sum_{i=1}^{50} \frac{\Delta E_i}{\kappa_i}, 
\end{equation}
where $\Delta E_j$ is the width of the spectral band, $\Delta E_i$ the width of the spectral interval (the sub-band), $\kappa_i$ is the sub-band absorption opacity, and $\kappa_j$ is the resulting opacity in the given spectral band, which is used in the solution of the radiation transfer equation.  

 \begin{table}
\caption{Mass fractions of chemical elements in the considered fusion ashes.
 \label{tab:comp} 
 }
{\footnotesize
\begin{center} 
\begin{tabular}{lcccc}
\hline \hline
 Element   & ash1& ash2 & ash3   & ash4  \\ 
\hline
 He	   &  0.834 &  0.209   &  0.185 & 0.072\\
 C	    & 0.164   &  2.46$\times 10^{-3}$&  1.9$\times 10^{-3}$ &  1.4$\times 10^{-5}$ \\
 Si        &  1.41$\times 10^{-3}$     &  0    &  1.23$\times 10^{-4}$ & 0 \\
 S	    &  1$\times 10^{-5}$  & 3.66$\times 10^{-4}$      &  1.23$\times 10^{-4}$ & 0 \\
 Ar        &     0             &  6$\times 10^{-4}$   &  1.82$\times 10^{-4}$ & 0 \\
 Ca       &      0             &  6.8$\times 10^{-2}$   &  0 & 3.5$\times 10^{-5}$ \\
 Ti        &        0          &   0.174  & 1.9$\times 10^{-3}$ & 8.9$\times 10^{-5}$ \\
  Cr       &       0           &  0.440   & 0.01 & 0.111 \\
 Fe        &      0            &   0.106  &  0.1  & 0.627 \\
 Ni        &       0           & 1.$\times 10^{-3}$   &  0.7 & 0.190 \\
\hline
\end{tabular}
\end{center}
}
\end{table}

\section{Results}
\label{sect:uncertainties}

The input parameters of the model are the chemical composition of the atmosphere, the surface gravity, and the effective temperature $T_{\rm eff}$.
We fix the surface gravity at the value $\log g = 14.3$ in cgs units for all the computed models.   
Some models with $\log g = 14.0$ were also computed for comparison. 
We note that a NS with commonly adopted parameters of mass $M = 1.5 M_\odot$ and radius $R = 12$\,km has $\log g = 14.24$. 
Increasing the mass to 2$M_\odot$ and decreasing the radius to 10~km yields $\log g \approx 14.6$, while a lower value of $\log g \approx 14.0$ corresponds to $M = 1.2M_\odot$ and $R = 14$\,km.
Thus, the expected range of surface gravity for NSs is $\log g \approx$14.0--14.6 in cgs units.

Four chemical compositions are chosen for our investigations as presented in Table\,\ref{tab:comp}.
They are taken from \citet{2018ApJ...863...53Y} and represent the chemical compositions for computations of X-ray bursts, ignited at three different column densities in the envelopes: $5\times10^8$ (y1n21), $1.5\times10^9$ (y1n22), and $5\times 10^9$\,g\,cm$^{-2}$ (y1n23), see their Fig.~3.
The chemical compositions {\sf ash1} and {\sf ash3} correspond to the values at the wind base of the models y1n21 and y1n23. 
{\sf ash2} corresponds to the chemical composition of the surface layers of the developed wind for the y1n22 model \citep[Fig.~9 in][]{2018ApJ...863...53Y}, while {\sf ash4} corresponds to the deepest layers of the same model. 
Thus, we have chemical composition patterns dominated by helium ({\sf ash1}), chromium ({\sf ash2}), nickel ({\sf ash3}), and iron ({\sf ash4}).
In addition, several models with pure iron composition were computed for comparison with previous results of \citet{Nattila.etal:15}. 

We also study models with an admixture of a plasma with solar H/He composition and reduced metal abundances. In these models, the mass fraction $X_{\rm ash}$ corresponds to one of three chemical compositions of the chosen
ash,  and the mass fraction  $1-X_{\rm ash}$ corresponds to the solar chemical composition with the hundred times reduced heavy element abundances ($A=0.01$).
The chemical composition of the atmosphere is assumed to be homogeneous throughout its depth.

For each chemical composition, a sequence of models with different relative luminosities $\ell=F/F_{\rm Edd}=$0.01, 0.03, 0.05, 0.07, 0.1, 0.15, 0.2, 0.3, 0.4, 0.5, 0.55, 0.6, 0.65, 0.7, 0.75, 0.8, 0.85, 0.9, 0.95, 0.98, 1.0, 1.02, 1.04, and 1.06 was calculated. 
For some chemical compositions, models with high luminosities could not be computed because the radiation force $g_{\rm rad}$ in some parts of the atmospheres became larger than the surface gravity~$g$. 
The effective temperatures are computed as $\sigma_{\rm SB}T^4_{\rm eff} =\ell F_{\rm Edd}$, where
\begin{equation} 
\label{eq2}
F_{\rm Edd} = \frac{c\,g}{\kappa_{\rm e}}, \qquad \kappa_{\rm e} = 0.2\,[1+(1-X_{\rm ash})\,X]~\rm {cm^2\,g^{-1}}.
\end{equation}
Here $\sigma_{\rm SB}$ is Stefan-Boltzmann constant, $\kappa_{\rm e}$ is the electron scattering Thomson opacity, and $X=0.7374$ is the solar hydrogen mass fraction. 
\footnote{Solar abundances of all chemical elements are taken from \citet{2009ARA&A..47..481A}.}

Radiation transfer is solved in a wide spectral interval, from $E_{\rm min} = (k_{\rm B}T_{\rm eff}/200)$\,keV to $E_{\rm max} = 50 \ k_{\rm B}T_{\rm eff}$\,keV. 
Every model atmosphere has 98 depth points logarithmically evenly spaced from depth $m\approx 3\times 10^{-7}$ to $\approx 3\times 10^{4}$\,g\,cm$^{-2}$. 
The depth of the bottom point is sufficiently large to satisfy the lower boundary condition used for the radiation transfer equation with the mean intensity approaching the Planck value, $J_{E} ({\rm bottom}) \approx B_E(T_{\rm bottom})$, see details in \citet{SPW12}. 
Spectral energy distributions of the escaping radiation for all models can be found in Zenodo (\href{https://zenodo.org/records/20035991}{https://zenodo.org/records/20035991}). 

\subsection{Physical properties of atmospheres enriched by heavy elements} 
\label{sec:hotmr}

\begin{figure} 
\begin{center}  
\includegraphics[width=  0.78\columnwidth]{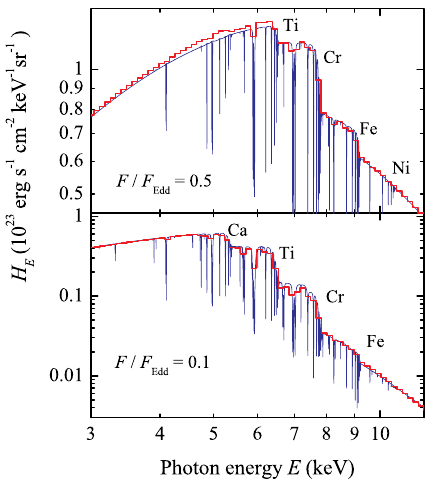}
\caption{\label{fig1} 
Spectra of emergent radiation (so called Eddington flux, $H_E=F_E/4\pi$) from NS model atmospheres with {\sf ash2} chemical composition. 
The upper panel corresponds to $\ell = 0.5$ and the lower panel to $\ell = 0.1$. 
The red and blue curves correspond to the models that include and omit Compton scattering, respectively.  
The photoionization edges of various chemical elements are marked.}
\end{center} 
\end{figure}

\begin{figure} 
\begin{center}  
\includegraphics[width=  0.78\columnwidth]{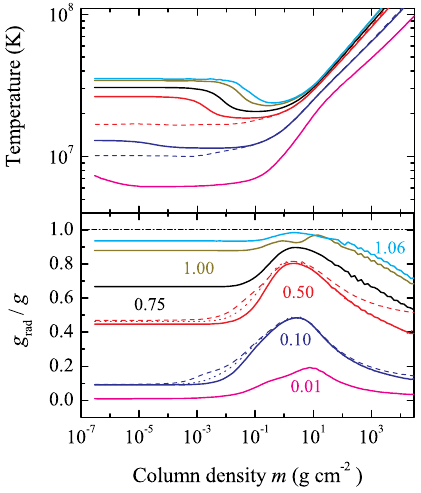}
\caption{\label{fig2} 
NS atmosphere structure for various relative luminosities $\ell$ (mix {\sf ash2},  $\log g$=14.3). 
Distribution of the temperature with depth $m$ is shown at the top panel, while the relative radiation pressure force is given at the bottom panel. 
The corresponding distributions for two models calculated in the Thomson scattering approximation are shown by dashed lines. 
The dotted lines show $g_{\rm rad}/g$ computed for the same two models using solution of the radiative transfer equation in all the sub-bands (without Compton effect). }
\end{center} 
\end{figure}

\begin{figure} 
\begin{center}  
\includegraphics[width=  0.78\columnwidth]{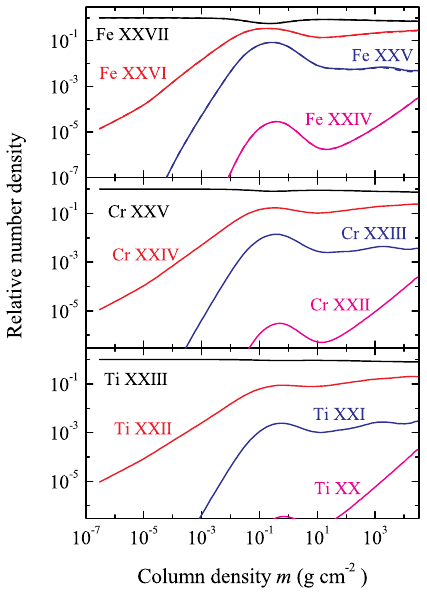}
\caption{\label{fig3} 
Distributions of the relative ion densities with depth for model atmospheres of chemical composition {\sf ash2}, $\ell = 0.1$, $\log g = 14.3$. 
The top, middle, and bottom panel correspond to iron, chromium, and titanium, respectively.}
\end{center} 
\end{figure}

Let us first consider the properties of models with a fiducial chemical composition {\sf ash2}. 
Under conditions of prevailing absorption opacity, Compton scattering is not very significant. 
Thus, we can verify our assumption about the averaging of the absorption opacity (see Eq.\,\eqref{eq:av}). 
For that we compute a few model atmospheres with and without consideration of the Compton recoil. 
The spectra of the model with $\ell = 0.1$ and 0.5 are shown in Fig.\,\ref{fig1}. 
The model spectra computed with Compton scattering (red curves) correspond well to the general appearance of the model spectra  calculated in the coherent electron (Thomson) scattering approximation (blue curves).  
In our approach, it is impossible to reproduce the full richness of the spectrum with spectral lines when accounting for Compton scattering, but it does not introduce errors into the overall energy distribution in the spectrum. 
We also tried direct averaging of the absorption opacity within the spectral bin with the same division into 50 sub-bins. 
However, in this case, the significance of opacity in spectral lines was greatly overestimated, and the overall appearance of the spectrum was severely distorted.

The temperature structures of models with the chemical composition {\sf ash2} are generally similar to the temperature structures of models  with solar chemical composition (see top panel of Fig.\,\ref{fig2}). 
The upper layers of the atmospheres demonstrate a chromosphere-like temperature increase due to the heating of electrons by the hard radiation from the inner atmospheric layers. 
However, the temperature distribution does not become flat at $\ell \approx 1$, as in solar chemical composition models, and a well-defined temperature minimum is retained even in models with luminosities close to the Eddington limit.
This is due to the significantly greater efficiency of thermal cooling of the atmosphere in the transition region between the inner optically thick layers of the atmosphere and its outer optically thin layers. This increased cooling efficiency is associated with increased absorption opacity caused by photoionization  and line transitions of heavy element ions.

The increase in opacity in these atmospheric layers also leads to a significant increase in the radiation pressure force in them (see lower panel of Fig.\,\ref{fig2}). 
Such a local increase in radiation pressure leads to the appearance of a ``levitating'' transition layer in the NS atmospheres  consisting of the products of thermonuclear combustion. 
In fact, the radiation pressure in this layer determines the value of the highest achievable bolometric flux of the atmosphere for a given chemical composition and surface gravity (see the upper line corresponding to $\ell = 1.06$ in the bottom panel of Fig.\,\ref{fig2}).

\begin{figure} 
\begin{center}  
\includegraphics[width=  0.78\columnwidth]{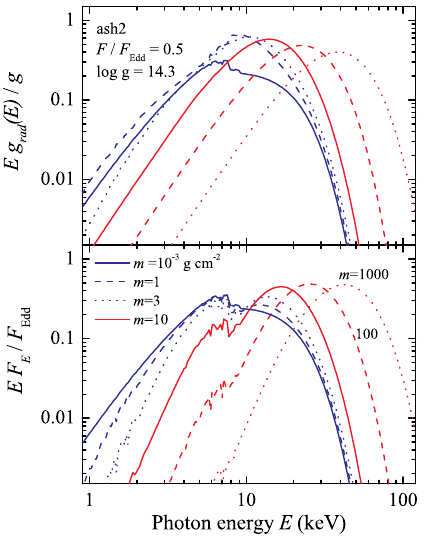}
\caption{\label{fig4} 
Variation of the spectral distribution of the   radiation pressure acceleration $E g_{\rm rad}(E)$ given by Eq.\,\eqref{grade} (top panel) and the net flux $E F_E$ (bottom panel) with atmospheric depth. 
The chemical composition and other parameters of the model atmosphere are shown in the top panel.}
\end{center} 
\end{figure}

We also verified the correctness of the opacity averaging procedure (see Eq.\,\eqref{eq:av}). 
In Fig.\,\ref{fig2} (lower panel), the dashed curves show the distribution of $g_{\rm rad}/g$ for two models with $\ell=0.1$ and 0.5, calculated in the coherent scattering approximation. 
Differences from solid curves at depths less than 1\,g\,cm$^{-2}$ are associated with differences in the temperature structure, and at greater depths with the fact that coherent scattering does not take into account the relativistic (Klein-Nishina) decrease in the scattering cross-section \citep[see][]{1983ApJ...267..315P, SPW12,2017ApJ...835..119P}. 
For each model, along with calculating the absorption opacity at each frequency point in spectral sub-bands, the radiation transfer equation was solved without the Compton effect taken into consideration. 
As a result, on the basis of these computations, the distributions of the relative radiative pressure force $g_{\rm rad}/g$ was calculated and shown at the bottom panel of Fig.\,\ref{fig2} by dotted curves which almost completely coincide with the solid curves. 
This means that the assumption used for averaging the absorption opacity (see Eq.\,\eqref{eq:av}) does not affect the radiation pressure force.

This increase in opacity and the corresponding hump in the distribution of the relative radiation pressure force $g_{\rm rad}/g$ are associated with the fact that in this transition layer of the atmosphere, the maximum relative number densities of hydrogen and helium-like ions of the most abundant heavy elements,  iron, chromium, and titanium, are achieved  (see Fig.\,\ref{fig3}).

The dependence of the ionization degree on the plasma density plays an important role in such number-density behavior. 
In the surface rarefied layers, heavy elements are almost completely ionized. 
In the deeper layers, the relative number densities of hydrogen and helium-like ions increase due to the increase in plasma density and reaches a maximum at depths 0.1-- 1\,g\,cm$^{-2}$. 
In even deeper layers, the relative number densities of these ions drop and stabilize due to pressure ionization.

Let us examine the cause of the levitating layer in more detail. 
Generally speaking, the force of radiation pressure is proportional to the product of the flux and the opacity. 
Therefore, at depths of 0.1--10\,g\,cm$^{-2}$, the condition must be satisfied that the maximum in the spectral flux distribution coincides with the maximum opacity. 
And maximum opacity, naturally, is achieved at energies near the absorption edges of the most abundant chemical elements.

We have illustrated the above reasoning with Fig.\,\ref{fig4}. It demonstrates how the spectral energy distribution of the flux $F_E$ changes with depth in the atmosphere (bottom panel). 
This distribution at large depth is close to the blackbody derivative with respect to the local temperature of the atmosphere at that depth, 
\be 
F_E \approx \frac{4\pi}{\sigma(E)+\kappa(E)}\, \frac{{\rm d}B_E}{{\rm d}T}\,\frac{{\rm d}T}{{\rm d}m}. 
\ee 
Here $\sigma(E)$ is angle-averaged electron scattering opacity accounting for Compton effect and  $\kappa(E)$ is the ``true'' absorption opacity (corresponding to  free-free and bound-free transitions).
Accordingly, the distribution maximum shifts from high photon energies at  larger depths towards lower energies at the surface layers.  
The influence of absorption edges on the spectral energy distribution becomes more noticeable closer to the surface. 
We note that at energies near the absorption edges of the most abundant chemical elements (Cr, Fe, Ti), around 10 keV, the flux is low at large depths, reaches a maximum at $m \approx 10$\,g\,cm$^{-2}$, and then at  lower depths falls again  to the values corresponding to the emergent flux.

The spectral energy distribution of the flux should be compared to the energy distribution of acceleration caused by the radiation pressure (shown in the top panel of Fig.\,\ref{fig4}) which can be defined as follows:
\be \label{grade}
g_{\rm rad}(E) = \frac{2\pi}{c} \int_{-1}^{+1} [\sigma(E,\mu)+\kappa(E)][I(E,\mu)-S(E,\mu)]\,\mu\,\mbox{d}\mu,
\ee
where $\mu$ is the cosine of the angle between the surface normal and the direction of radiation propagation,  $I(E,\mu)$ and $S(E,\mu)$ are the specific intensity and the source function, respectively (see  Sect.~2.1 of \citealt{SPW12} for detailed description).
In general, the behavior of $E g_{\rm rad}(E)$ with depth corresponds to the behavior of the flux  $EF_E$. 
However, due to the influence of opacity, the integral of $g_{\rm rad}(E)$ over energy is not constant with the depth. 
At large depths, it is relatively small due to the decrease in the electron scattering cross section at high energies (Klein-Nishina effect), and the true opacity is negligible.
As the depth decreases and the flux maximum shifts toward lower energies, the contribution of the opacity caused by photoionization from the ground level of hydrogen-like ions increases. 
This contribution reaches maximum at depths 1--10\,g\,cm$^{-2}$, where the maximum in the flux spectral energy distribution approaches dominating photoionization edges and the relative ionization degree   becomes minimal. 
Close to the surface ($m < 10^{-3}-10^{-4}$\,g\,cm$^{-2}$), the contribution of true opacity again becomes negligible due to the decrease in plasma density. 
In these outer atmospheric layers the spectral distribution of the radiative acceleration coincides with the spectral distribution of the flux, since there the electron scattering opacity is practically independent of the photon energy. 

\begin{figure} 
\begin{center}  
\includegraphics[width=  0.78\columnwidth]{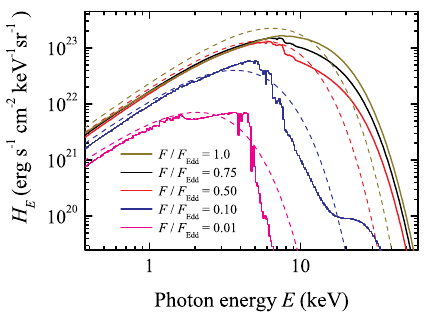}
\caption{\label{fig5} 
Emergent spectra for the models presented in Fig.\,\ref{fig2}. 
The model spectra are shown with the solid curves, while the corresponding blackbody spectra for a few models of the same effective temperature are shown with the dashed curves.}
\end{center} 
\end{figure}

\begin{figure} 
\begin{center}  
\includegraphics[width=  0.78\columnwidth]{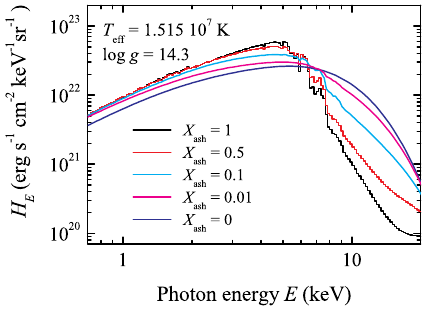}
\caption{\label{fig6} 
Emergent spectra of the models composed of a plasma with {\sf ash2} composition ($X_{\rm ash}$ mass fraction) mixed with solar abundance 
plasma ($A=0.01$, $1-X_{\rm ash}$ mass fraction). The effective temperature is fixed and corresponds to $\ell=0.1$ for $X_{\rm ash}=1$.
}
\end{center} 
\end{figure}

\begin{figure} 
\begin{center}  
\includegraphics[width=  0.78\columnwidth]{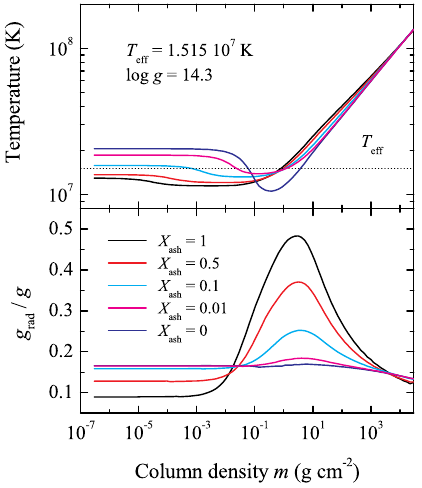}
\caption{\label{fig7} 
Temperature structures (top panel) and $g_{\rm rad}/g$ distributions (bottom panel) for the models with the mixed chemical compositions (see Fig.\,\ref{fig6}).
}
\end{center} 
\end{figure}

The spectra of model atmospheres, whose temperature structures are shown in Fig.\,\ref{fig2}, are shown in Fig.\,\ref{fig5} (except for the model $\ell = 1.06$).
The spectra of models with low relative bolometric flux, up to $\ell = 0.5$ , are determined by the photoionization edges  from the ground states of hydrogen- and helium-like 
ions and absorption lines of these ions. In fact, the Compton effect can be disregarded when modeling atmospheres with this chemical composition and low relative luminosities
(see also Fig.\,\ref{fig1}). It should be noted that the absorption edges are also significant in the spectra of models with higher relative bolometric flux, disappearing almost 
completely only when approaching the Eddington bolometric flux. Compared to the blackbody, the model spectra with the same effective temperature exhibit an increased radiation flux at energies below the main absorption edge, and significantly lower at energies above it. 

As expected, the influence of absorption due to photoionization of heavy element ions and their spectral lines decreases with the decrease of the mass fraction 
of thermonuclear ashes ($X_{\rm ash}$) in the composition of the atmospheric plasma (see Fig.\,\ref{fig6}). Here we consider the mix of {\sf ash2} composition 
with the plasma of solar chemical composition with a hundred times reduced heavy element abundances ($A$=0.01). All the models have the same effective 
temperature corresponding to $\ell = 0.1$ for $X_{\rm ash}$=1. For other models the relative bolometric flux increases up to $\ell \approx 0.17$ at 
$X_{\rm ash}$=0 according to Eq.\,\eqref{eq2}. The temperature structures and distributions of the relative radiation force $g_{\rm rad}/g$ for 
these models are shown in Fig.\,\ref{fig7}. It is evident that the significance of the levitating transition layer decreases with a decrease in the mass fraction of the ashes (see bottom panel of Fig.\,\ref{fig7}).

\begin{figure} 
\begin{center}  
\includegraphics[width=  0.78\columnwidth]{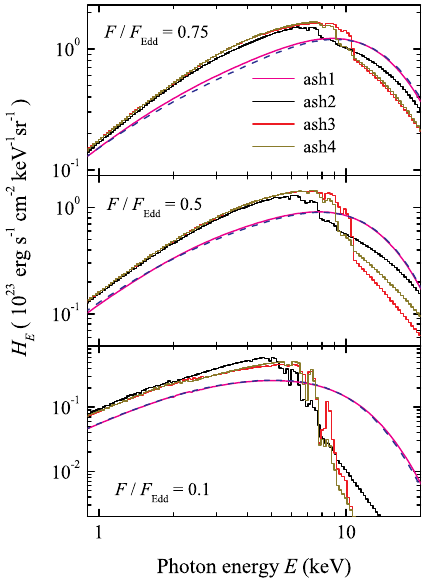}
\caption{\label{fig8} 
Emergent spectra of models consisting of various mixtures of  thermonuclear combustion products ({\sf ash1}, {\sf ash2}, {\sf ash3}, and {\sf ash4} (see Table\,\ref{tab:comp}), and various relative bolometric fluxes, $\ell = 0.75$ (top panel), 0.5 (middle panel), and 0.1 (bottom panel). 
The emergent spectra for pure helium models are shown with dashed blue curves.}
\end{center} 
\end{figure}

\begin{figure} 
\begin{center}  
\includegraphics[width=  0.78\columnwidth]{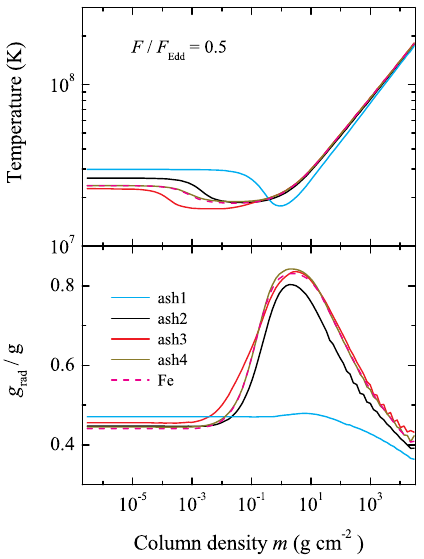}
\caption{\label{fig9} 
Temperature structures (top panel) and $g_{\rm rad}/g$ distributions (bottom panel) for the models with  $\log g = 14.3$, $F/F_{\rm Edd}=0.5$, and 
different chemical compositions.
}
\end{center} 
\end{figure}

The appearance of model spectra for atmospheres consisting of thermonuclear combustion products is completely determined by the composition of the ashes, especially at low relative luminosities (see Fig.\,\ref{fig8}). 
The most significant absorption edge in the spectra shifts toward higher energies during the transition from composition {\sf ash2} to composition {\sf ash4}, and further to composition {\sf ash3},  since chromium dominates in the first composition, iron in the second composition, and nickel is the most abundant element in the third one (see Table\,\ref{tab:comp}). 
This property of spectra can be traced up to the models with the highest relative bolometric flux. 
In composition {\sf ash1}, helium is the most abundant element, and as a result the emergent spectra of models with this chemical composition are practically indistinguishable from the spectra of pure helium model atmospheres (compare solid magenta and dashed blue curves in Fig.\,\ref{fig8}).

The chemical composition of the atmosphere affects its temperature structure and the significance of radiation pressure  (see Fig.\,\ref{fig9}). 
The temperature structure and distribution of the relative radiation pressure force $g_{\rm rad}/g$ across the atmosphere, calculated for an atmosphere consisting of pure iron, are also shown. 
They are virtually identical to the results for {\sf ash4}. 
The significance  of the levitating layer increases with the transition from composition {\sf ash2} to {\sf ash3} and to {\sf ash4} (see an example in the bottom panel of Fig.\,\ref{fig9}).  
In high bolometric flux models calculated with compositions {\sf ash3} and {\sf ash4} (and pure iron), this layer determines the highest possible bolometric flux. 
The force of radiation pressure exceeds the force of gravity already at relative luminosities of 0.77--0.78. 
Therefore, models $\ell=0.75$ are the models with the highest bolometric flux calculated for these chemical compositions.

Here we did not consider atmospheres with chemical compositions dominated by elements heavier than nickel. However, we expect that their properties should generally not differ significantly from iron- or nickel-dominated models. 
It is clear from the comparison of the spectra of iron- and nickel-dominated models (see Fig.~\ref{fig8}) that for zinc-dominated model atmospheres the main absorption edge will shift to 12.4\,keV (the ionization energy of hydrogen-like zinc), but otherwise the spectra will be rather similar.

\subsection{Spectral evolution of the models enriched by heavy elements}
\label{sect:grids}

The dependence of emergent spectra of model atmospheres of hot NSs on the relative flux is important for the study of X-ray bursts in LMXBs.
Of particular interest is the degree to which the model spectrum deviates from the blackbody of a temperature equal to the effective temperature of the model. 
If we fit the model spectra by the spectra of a diluted blackbody $w\,\pi\,B_E (f_{\rm c} T_{\rm eff})$, then these deviations can be conveniently characterized by the fit parameters, the spectral dilution factor $w$ and the color correction factor $f_{\rm c}$. 
If a thermonuclear burst occurs during a hard spectral persistent  state of the source, then the best-fit blackbody normalization $K=(R_{\rm BB}/d)^2$ at the brightness decline stage changes with decreasing flux in exactly the same way as the dilution factor $w$ behaves with decreasing relative flux $F/F_{\rm Edd}$. 
This fact allows us to obtain an estimate of the NS radius \citep[see, e.g.,][]{Suleimanov.etal:17}.

\begin{figure} 
\begin{center}  
\includegraphics[width=  0.78\columnwidth]{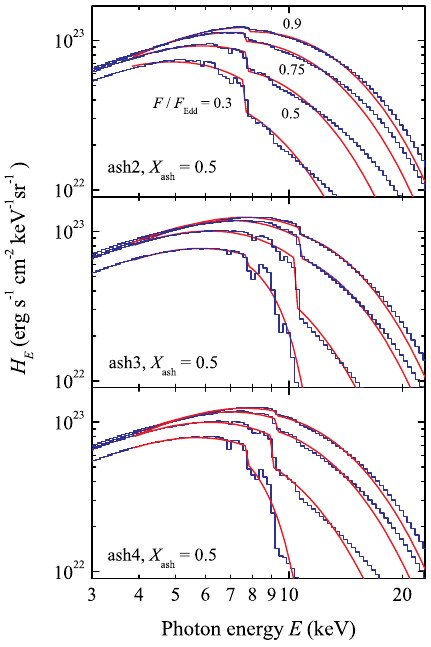}
\caption{\label{fig10} 
Best fits of model spectra  by  the blackbody with the edge. 
The relative fluxes are $l=0.3$, 0.5, 0.75, and 0.9. 
The top, middle, and bottom panels correspond to  {\sf ash2}, {\sf ash3}, and {\sf ash4} chemical compositions half-mixed with solar-abundance plasma of $A=0.01$. 
The model spectra are the blue histograms, while the best-fits are shown with red curves.}
\end{center} 
\end{figure}

The presence of noticeable absorption edges in the model spectra makes their fitting by diluted blackbodies inaccurate.
We had previously ignored this \citep{SPW12, Nattila.etal:15}, but in this paper the spectra of model atmospheres enriched in fusion products 
have such significant absorption edges that we decided to use a more complex fitting function. 
This is a five-parameter analytical function, based on the diluted blackbody spectrum with one absorption edge: 
\begin{equation} 
\label{eq:FwBE}
    F^{\rm fit}_E =  
   \left\{ 
   \begin{array}{ll}
   w \,\pi B_E(f_{\rm c} T_{\rm eff}),  & \textrm{if}\ E \leq E_{\rm th}, \\ 
    w \,\pi B_E(f_{\rm c} T_{\rm eff})\   \exp(-\tau_{\rm th}\  [E/E_{\rm th}]^{-p}), &  \textrm{if}\ E > E_{\rm th}, 
    \end{array} \right.         
\end{equation}
where $E_{\rm th}$ is the threshold energy of the most important absorption edge in the spectrum, $\tau_{\rm th}$ is the optical depth at the threshold, and $p$ is the absorption exponent.

In fact, this function is a combination of the diluted blackbody and the \textsc{xspec} \citep{Arnaud1996} function \texttt{edge}.\footnote{https://heasarc.gsfc.nasa.gov/docs/software/xspec/manual/\\node252.html} 
The difference is that the absorption exponent is a free parameter that allows one to better describe the shape of the spectrum at photon energies exceeding the energy of the  absorption threshold, and it is not fixed at $p=3$ as in the original function \texttt{edge}.

\begin{figure} 
\begin{center}  
\includegraphics[width=  0.78\columnwidth]{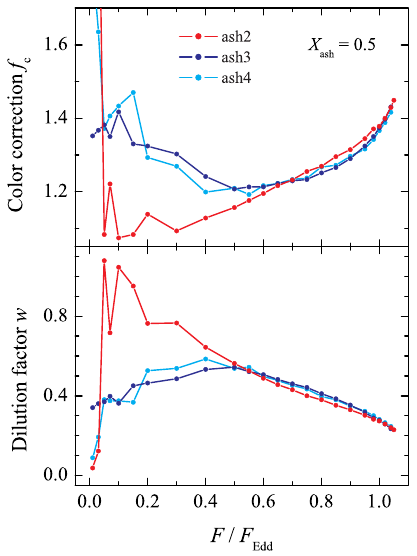}
\caption{\label{fig11} 
Dependence of the fit parameters on the relative flux for models with the same chemical composition as in Fig.\,\ref{fig10}. 
The color correction factor $f_{\rm c}$ is at the top panel and the dilution factor $w$ at the bottom panel). }
\end{center} 
\end{figure}

\begin{figure} 
\begin{center}  
\includegraphics[width=  0.78\columnwidth]{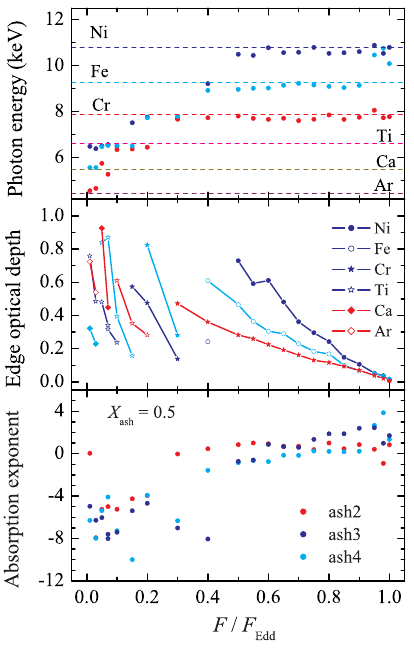}
\caption{\label{fig12} 
Dependence of the absorption edge fit parameters on the relative flux for models with the same chemical composition as in Fig.\,\ref{fig10}. 
Dependence of the threshold energy $E_{\rm th}$ (top panel), the optical depth at the threshold $\tau_{\rm th}$ (middle panel), and the absorption exponent $p$ (bottom panel) are shown. The dashed lines in the top panel show the threshold energies for hydrogen-like ions of the main chemical  elements. 
The optical thickness of the various absorption edges are shown by the different symbols in the middle panel. }
\end{center} 
\end{figure}

As in the previous works \citep[see, e.g.,][]{SPW11}, the fitting was carried out in the sensitivity band of the PCA/{\it RXTE} detector, 3--20 keV, shifted to the blue part of the spectrum in order to account for  the assumed gravitational redshift on the NS surface. 
The redshift was calculated depending on the adopted value of gravity and a fixed mass of the NS, $M = 1.4\,M_\odot$.

Examples of fitting spectra of model atmospheres with different chemical compositions and different relative fluxes are shown in Fig.\,\ref{fig10}. 
The main absorption edge in the spectra of high relative bolometric flux is determined by the chemical composition of the atmosphere, i.e. these are absorption edges from the ground level of hydrogen-like ions of nickel ({\sf ash3}, $X_{\rm ash} = 0.5$), iron ({\sf ash4}, $X_{\rm ash} = 0.5$), or chromium ({\sf ash2}, $X_{\rm ash} = 0.5$).  
The dependencies of the fitting parameters on the relative flux for the spectra of model atmospheres with the chemical composition presented above are shown in Figs.\,\ref{fig11} and \ref{fig12}.  
Interestingly, the basic fitting parameters  $w$ and $f_{\rm c}$ nearly coincide for all three chemical compositions if the relative flux exceeds 0.5 (see Fig.\,\ref{fig11}). 
However, at lower relative fluxes, the fitting parameters for chromium-dominated atmospheres begin to differ significantly from those for iron- and nickel-dominated atmospheres. 
For spectra of atmospheres with these two chemical compositions, the fitting parameters remain similar at all relative fluxes.

We note that with decreasing relative flux and shifting of the spectrum maximum towards lower energies, the photoionization edges from the ground levels of ions with lower ionization potentials become successively dominating (see upper panel of Fig.\,\ref{fig12}). 
The optical depth near the edge $\tau_{\rm th}$ increases with decreasing relative flux, as long as the main edge remains the same (see middle panel of Fig.\,\ref{fig12}). 
When the absorption edge in the fit changes, the optical thickness again becomes relatively small and increases with a decrease in the relative flux until the next change in the main edge in the fit. 

The absorption exponent $p$ varies over a very wide range (see bottom panel of Fig.\,\ref{fig12}). 
At high relative fluxes, it is mostly positive and fluctuates around zero.
However, at low relative fluxes, when the spectra become very complex with several absorption edges with a general significant drop in flux towards high energies, the absorption component becomes negative and fluctuates around $p \approx -6$. 
Such a large absolute value of $p$ and its great variability arise as a result of attempts to describe complex model spectra (see, e.g., the two bottom panels of Fig.\,\ref{fig10}, the spectra corresponding to $l = 0.3$).

\begin{figure} 
\begin{center}  
\includegraphics[width=  0.78\columnwidth]{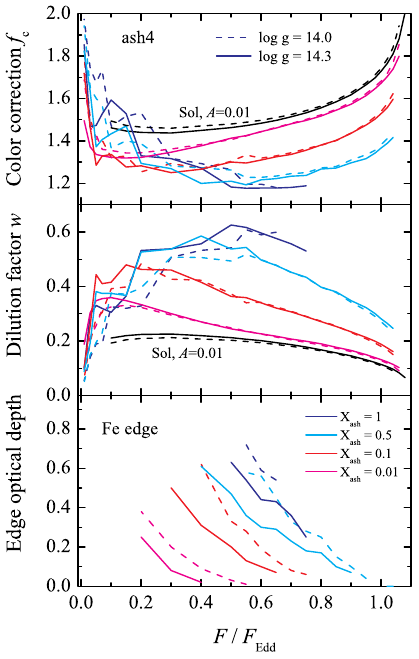}
\caption{\label{fig13} 
Dependence of the fit parameters on the relative flux for models with chemical composition {\sf ash4} and various values of $X_{\rm ash}$. 
The fit parameters obtained for $\log g = 14.3$ (solid curves) and $\log g =14.0$ (dashed curves) are shown.
For comparison, the fit parameters for models with $X_{\rm ash}=0$ are also shown with black curves in the two upper panels. 
Only the optical depth at the iron edge is shown (bottom panel). }
\end{center} 
\end{figure}

\begin{figure} 
\begin{center}  
\includegraphics[width=  0.78\columnwidth]{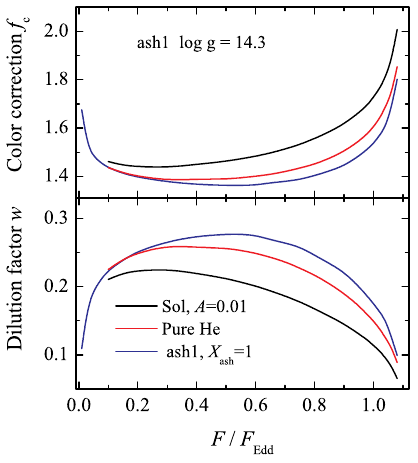}
\caption{\label{fig14} 
Dependence of the fit parameters on the relative bolometric flux for various chemical composition. 
The blue, red, and black curves correspond to models {\sf ash1},  pure helium, and solar H/He mix with sub-solar heavy elements abundances ($A=0.01$), respectively. }
\end{center} 
\end{figure}

Clearly, the shape of the model spectra, and therefore the spectral fitting parameters, depend on the proportion of thermonuclear ashes  in the atmosphere. 
An example of how the parameters change depending on the ash contribution to the chemical composition of the atmosphere and the relative flux is shown in Fig.\,\ref{fig13}. 
The chemical composition {\sf ash4} was chosen and calculations were performed for two values of gravity, $\log g = 14.3$ and 14.0.
The  color correction $f_{\rm c}$ decreases with increasing proportion of the ashes in the chemical composition of the atmosphere from 1.6 at $X_{\rm ash} = 0$ to 1.3  at $X_{\rm ash} = 0.5$ for the fixed relative flux $l = 0.9$ (see top panel of Fig.\,\ref{fig13}). 
Accordingly, the dilution factor $w$ increases with the increase in the proportion of ashes from 0.15 at $X_{\rm ash} = 0$ to 0.35 at $X_{\rm ash} = 0.5$ for the same fixed relative flux (see middle panel of Fig.\,\ref{fig13}). 

Note that the differences for $f_{\rm c}$ and $w$ for models with different gravities are small at large relative fluxes ($l> 0.5$). 
However, the difference is noticeable in the optical depth at the iron edge (see bottom panel of Fig.\,\ref{fig13}).  
The optical depth is higher for low gravity at the same relative flux. 
The reason is the lower effective temperature of the reduced gravity models at a fixed relative flux (see Eq.\,\ref{eq2}). 
Naturally, the iron ionization degree  decreases and the photoionization absorption edge becomes more significant.
Of course, the optical depth also depends on the abundance of ashes in the atmosphere. 
The lower the ash fraction, the lower the relative fluxes at which the absorption edge appears. 
Also the optical depth is smaller for lower ash fraction at a fixed $l$.

Hydrostatic model atmospheres consisting only of iron- ({\sf ash4}) or nickel-dominated ({\sf ash3}) ashes exist only up to the relative flux of the order of 0.65--0.75.
At higher luminosities, the radiation pressure force in the levitating  layer becomes greater than gravity, $g_{\rm rad} > g$, and the models become unstable. 
Moreover, spectra for model atmospheres composed solely of ashes are strongly distorted by absorption edges and spectral lines, so fitting them with the simple function (Eq. \ref{eq:FwBE}) does not always yield satisfactory results.

The spectra of model atmospheres consisting of helium-dominated ash material ({\sf ash1}) do not contain any noticeable absorption edges and can be fitted with a diluted blackbody. 
The results of such fitting are presented in Fig.\,\ref{fig14}. 
It is evident that the contribution of heavy elements leads to differences from the model spectra for pure helium.

In previous work on NS atmospheres enriched in heavy elements \citep{Nattila.etal:15}, pure iron atmospheres were considered. 
A comparison of such atmospheres with the iron-dominated atmospheres presented here ({\sf ash4}) is presented in Appendix~\ref{app:iron}.
The dependencies of the fitting parameters on the relative flux for all considered chemical compositions are presented in the tables and figures of  Appendix~\ref{app:fits}.

\section{Discussion}
\label{sect:discussion}

\subsection{Hydrostatic equilibrium and the role of radiation pressure acceleration}
\label{sec:radpress}

In this paper, we examine the properties of hot NS model atmospheres composed of chemical elements representative of thermonuclear ashes. 
Four distinct mixtures are analyzed, each dominated by helium, chromium, iron, or nickel. 
For each composition, we account for all opacity sources produced by ions of all elements present. 
They comprise photoionization transitions originating not only from ground levels but also from excited levels of the ions
In addition, about 5000 bound-bound transitions between ion levels (spectral lines) were included. 
These are the first calculations of NS model atmospheres taking into account such a large number of spectral lines.
Since models close to the Eddington limit were considered, Compton scattering of photons on electrons was also taken into account. 
A special opacity-averaging method was developed to simultaneously account for the Compton effect and the contribution of spectral lines to opacity.

An intriguing structural feature appears in the calculated model atmospheres -- namely, a layer where the ratio of the radiation pressure force to gravity, $g_{\rm rad}/g$, becomes markedly elevated. 
This region resides at column densities of a few grams per square centimeter, in the transition zone between the optically thick and optically thin layers.
This region exhibits a local minimum in the ionization degree of the heavy elements composing the atmosphere. 
Consequently, the relative number densities of helium-like and hydrogen-like ions of these elements reach their maximum there.

Because the radiation pressure force is at its maximum in the transition layer, the hydrostatic stability of these model atmospheres is controlled by this levitating layer of elevated radiative acceleration. 
Models for which the ratio $g_{\rm rad}/g$ in this region surpasses unity are inherently unstable.
As a result, models composed of mixtures dominated by iron or nickel become unstable already at relative fluxes $F/F_{\rm Edd} \approx 0.75$.  
Chromium-dominated models remain stable up to much higher fluxes, but in all cases, the origin of the instability is the levitating layer. 
In helium-dominated models, no levitating layer forms. 

\subsection{Emergent spectra}
\label{sec:spectra}

At atmospheric temperatures typical of X-ray bursters during flashes, heavy elements near the iron peak are not fully ionized. 
As a result, the dominant features in the emergent spectra of models of such atmospheres are absorption edges at energies of 5--10~keV, depending on the effective temperature and chemical composition. 
Therefore, the emergent spectra of atmospheres composed of heavy elements cannot be accurately fitted with a simple diluted blackbody and require the inclusion of an absorption edge.
However, the standard \textsc{xspec} \texttt{edge} procedure is not suitable, as it assumes that absorption above the photoionization threshold $E_{\rm th}$ arises from an optically thin layer, with the absorption coefficient decreasing with energy as $(E/E_{\rm th})^{-3}$. 
However, in our case, this exponent should not describe the energy dependence of the absorption coefficient, but rather the shape of the emergent spectrum above the absorption edge. 
Therefore, we treated the absorption exponent as an additional free parameter in the fit.

We fitted all calculated model-atmosphere spectra using a combination of a diluted blackbody and a modified edge procedure.  
It should be noted that we calculated not only models with chemical compositions corresponding to thermonuclear ashes, but also mixtures with plasma of solar composition, in which the abundances of heavy elements are reduced by a factor of 100 ($A=0.01$), with the ashes contributing relative mass fractions $X_{\rm ash}$ of 0.5, 0.1, and 0.01.

Overall, the results are similar to those reported by \citet{Nattila.etal:15} for atmosphere models with initially solar composition, but with enhanced heavy-element abundances. Specifically, increasing the mass fraction of ashes leads to a lower color-correction factor $f_{\rm c}$ and a higher dilution factor $w$. 
The  $F/F_{\rm Edd}$--$f_{\rm c}$ and  $F/F_{\rm Edd}$--$w$ dependencies obtained for pure ash mixtures dominated by iron and nickel differ qualitatively from those of other chemical compositions. 
In addition to the instability of models with relative fluxes above 0.75, these dependencies are nearly flat for relative fluxes above 0.4--0.5.  
It should also be noted that the resulting parameters differ depending on the fitting method used--whether using only a diluted blackbody or including an absorption edge.
In the latter case, the color-correction factor is larger and the dilution factor smaller than when fitting with a diluted blackbody alone. 
This should be kept in mind when comparing the calculated fitting parameters with observed X-ray burster spectra, which are often approximated by a blackbody without accounting for absorption edges.

Absorption edges in the spectra become noticeable when the relative flux falls below a certain threshold. 
The smaller the ash mass fraction, the lower the relative flux at which these edges become significant.
Since the fitting function includes only a single edge, its effect shifts sequentially from the most highly charged hydrogen-like ions to the less charged ions as the relative flux decreases. 
The optical depth of the edge near the ionization threshold, $\tau_{\rm th}$, increases as the relative flux decreases, as long as it corresponds to the same ionization threshold.  
However, it decreases when the fitted edge shifts to a different photoionization threshold.
As a result, $\tau_{\rm th}$ never exceeds unity in the calculated models.  
This is because, in addition to the dominant element, thermonuclear ashes contain substantial amounts of other heavy elements. 
To achieve a fitted $\tau_{\rm th}$ significantly greater than one, the atmosphere must be chemically pure, composed of only a single heavy element (see Appendix~\ref{app:iron}).

\subsection{Comparison with observations}
\label{sec:obs}

Let us now compare our modeling results with previously published observations of relatively long X-ray bursts that show clear evidence for the presence of thermonuclear ashes in NS atmospheres. 
The first example is an X-ray burst in the system HETE,J1900.1$-$2455, which occurred on 2005 July 21 and was observed by {\it RXTE} \citep{2017MNRAS.464L...6K}. 
By fitting the burst spectra with a blackbody, a dependence of $F_{\rm BB}$ on $K^{-1/4}$ was obtained for the cooling phase, which corresponds to the theoretical $F/F_{\rm Edd}$--$f_{\rm c}$ relation. 
Here, $K$ is the normalization and $F_{\rm BB}$ is the bolometric flux derived from the fit. 
The resulting dependence is non-monotonic. 
When the flux decreases by roughly a factor of two, $K^{-1/4}$ increases abruptly by about 20\% (see Fig.\,4 in the cited paper).
At $F/F_{\rm Edd}\approx 0.5$, the color-correction factor $f_{\rm c}$ for atmospheres composed of thermonuclear ashes is lower by roughly the same amount compared to models with solar composition (see, e.g., Fig.~\ref{fig13}).
Therefore, we concluded in \citet{2017MNRAS.464L...6K} that at this point the atmospheric composition changed from being dominated by thermonuclear ashes to reflecting the composition of the accreted matter due to the onset of accretion.

Two features of this burst are noteworthy. 
First, the slope of the curve at high luminosities, when the atmosphere was ash-dominated, is steeper than in the theoretical models, both those previously calculated by \citet{Nattila.etal:15} and the ones presented here.
This was interpreted as a gradual increase in the thermonuclear ash fraction in the atmosphere, reaching a maximum just before the onset of accretion. 
This change in chemical composition can occur for two reasons. 
Either the NS surface loses material through a wind, exposing deeper layers richer in thermonuclear reaction products, or convection in the atmosphere brings thermonuclear ashes to the surface. 
The most likely mechanism for mass loss is a line-driven wind generated by radiation pressure in spectral lines \citep[see, e.g.,][]{2008A&ARv..16..209P}. 
However, in the models presented here, line-driven radiative pressure is negligible and insufficient to launch a wind. 
This is because the hydrogen- and helium-like ions dominating these models do not possess a sufficient number of bound-bound transitions near the peak of the emergent spectral energy distribution. 
However, the situation could change if the ashes contain elements heavier than nickel, such as zinc, germanium, selenium, or krypton. 
These heavier elements may be less ionized and possess three to five bound electrons.
Such ions generate a large number of spectral lines \citep[cf. hot white dwarf atmospheres; see, e.g.,][]{2024A&A...688A..39S}, which could potentially drive the formation of a line-driven wind. 

Regarding the second possibility --dredge-up of heavy elements to the surface by convection-- the chemically homogeneous atmosphere models presented here are formally stable according to the Schwarzschild criterion. 
Specifically, the logarithmic pressure gradient with respect to temperature is everywhere smaller than the adiabatic gradient for a radiation-pressure-dominated plasma, $d \log P/d \log T < 0.25$. 
The partial ionization of heavy elements in the models may reduce the adiabatic gradient, although this effect has not been investigated. 
However, the fraction of incompletely ionized atoms does not exceed 10\% even in models with low relative flux, and only within the levitating layer. 
Thus, if a relatively thin convective layer forms there, it is unlikely to have any significant global effect on the atmosphere’s properties.

The second notable feature of the burst is the appearance of a clear absorption edge in the spectrum just before the onset of accretion, when the abundance of heavy elements in the atmosphere was presumably at its maximum. 
The edge was observed at an energy of approximately 7.6~keV and had a relatively small optical depth near the threshold, $\tau_{\rm th} \approx 0.6$. 
This value suggests that the ash in the atmosphere was not chemically pure, but rather a mixture corresponding to either {\sf ash4} (if the observed jump is due to the iron ionization threshold) or {\sf ash3} (if it corresponds to the nickel ionization threshold) considered in our models.

The second burst of interest  occurred on 2002 April 30 in the system GRS\,1747$-$312 and was also observed by {\it RXTE} 
\citep{2018ApJ...866...53L}. 
Its properties differ from those of the previously discussed burst: there is no indication of accretion onset, and the observed $F_{\rm BB}$--$K$ curve is best reproduced by pure-iron models calculated by \citet{Nattila.etal:15}.
However, our calculations indicate that atmospheres composed of pure iron, or dominated solely by iron or nickel, cannot exist near the Eddington limit.
This suggests that the NS atmosphere during this burst was most likely not composed of pure ash, but contained a significant admixture of solar-composition plasma, with $X_{\rm ash} \approx$0.5--0.8. 
The absorption edge in the spectra of this burst appears at luminosities below roughly half of the maximum and is observed at a slightly higher energy than in the previously discussed burst, around 8~keV.
However, the optical depth at the threshold is significantly higher,  $\tau_{\rm th} \approx 2-3$ or even larger. 
This suggests that the thermonuclear ash in the NS atmosphere during this burst was nearly chemically homogeneous, with one heavy element globally dominant. 
This conclusion follows from the fact that  $\tau_{\rm th}$ cannot exceed unity if the atmosphere contains multiple heavy elements (see Fig.~\ref{fig:a2}).
 
\citet{2018ApJ...866...53L} performed spectral fits of the burst using both a blackbody and a blackbody with an absorption edge (see their Fig.~2). 
The results indicate that the fit including the edge yields a lower normalization $K$ and a higher temperature compared to a blackbody fit without the edge. 
These findings are in complete agreement with the fitting results for our model spectra, both in terms of the color-correction factors $f_{\rm c}$ and the dilution factors $w$
 (see Figs.~\ref{fig:a4} and \ref{fig:B1}–\ref{fig:B3}).

\section{Conclusions}
\label{sect:conclusions}

In the work presented here, we investigated the properties of the hot NS atmospheres enriched in products of thermonuclear reactions (thermonuclear ashes).
To this end, we developed a method for calculating model atmospheres taking into account both a large number of spectral lines and the Compton effect.
Specifically, a procedure for averaging the rapidly changing line opacity over wide spectral bins was proposed, and used to solve the radiative transfer equation accounting for Compton scattering.

A total of four possible mixtures were considered to describe the thermonuclear ashes. 
They can be characterized by the dominant element in their composition, namely, helium, chromium, iron, or nickel, although significant amounts of other chemical elements are also present in each mixture. 
Almost all models were calculated at fixed gravity ($\log g = 14.3$) and for a wide range of relative bolometric fluxes $l$ from 0.01 and up to the highest possible flux for a given chemical composition and $\log g$. 
For most of the mixtures considered, the maximum escaping flux exceeds $F_{\rm Edd}$, although for mixtures dominated by iron and nickel, the limiting flux is close to 0.75$F_{\rm Edd}$. 

Unlike models of atmospheres dominated by light elements, where the limiting bolometric flux is determined by photon scattering on electrons in the upper layers of the atmosphere, this is not the case in the models considered. 
In models of hot NS atmospheres dominated by heavy elements, there exists a levitating  layer with an increased ratio of the radiation pressure force to the force of gravity, $g_{\rm rad}/g$. 
It is located in the transition zone between the optically thick and optically thin layers of the atmosphere, and the $g_{\rm rad}/g$ ratio hill owes its existence to the radiation pressure in the photoionizing continua of heavy element ions. 
It is the excess of $g_{\rm rad}$ over $g$ in that layer which determines the value of the maximum bolometric flux in the sequence of models of a given chemical composition.

In mixtures dominated by chromium or heavier elements, the absorption edges are noticeable in the model spectra.
Therefore, we fitted them with a diluted blackbody with an additional  absorption edge.
Typically, model spectra have several absorption jumps, so the lowest-energy edge was chosen as the primary absorption threshold. 
The model spectrum at energies above this absorption edge was fitted by varying an additional parameter in the edge description, the absorption exponent. 

We investigated how the fitting parameters -- the color correction $f_{\rm c}$, the dilution factor $w$, and the absorption edge optical depth at the threshold $\tau_{\rm th}$-- depend on the relative bolometric flux of the atmosphere and its chemical composition. 
We have shown that the more the atmosphere is enriched with thermonuclear ashes, the smaller the color correction and the larger the dilution factor. 
These dependencies are practically flat at relative bolometric fluxes above 0.5 for iron- or nickel-dominated atmospheres.

The optical thickness of the absorption edge increases with decreasing relative bolometric flux of the atmosphere, but will not exceed unity if the mixture contains significant amounts of several heavy elements, such as titanium, chromium, iron, and nickel. 
The reason is that as the bolometric flux, and therefore the effective temperature of the atmosphere, decreases, the role of the main absorption edge continually shifts to ions of ever lighter elements. 
The optical thickness of the absorption edge exceeds unity only in chemically homogeneous atmospheres consisting of one of the aforementioned species.

A qualitative comparison of the calculated model spectra with the results of fitting two unusual X-ray bursts from the LMXBs HETE~J1900.1$-$2455 and GRS~1747$-$312 showed that the models previously considered may not be sufficient to explain them. 
In particular, comparison with our new models confirmed the conclusion made by \citet{2017MNRAS.464L...6K} that the relative abundance of thermonuclear ashes in the atmosphere changes rapidly during the considered X-burst in HETE~J1900.1$-$2455. 
Therefore, it is necessary to consider chemically stratified model atmospheres.
We expect that in such atmospheres, the plasma opacity at high photon energies ($E > 10$\,keV) will increase more rapidly with depth. 
This will lead to a decrease in the emergent radiation flux at these energies. 
Overall, the spectrum of the emergent radiation will be closer to the spectrum of a blackbody with a temperature equal to the effective temperature resulting in a decrease of the color correction factor.

The contribution of elements heavier than nickel to the chemical composition of the ashes may also be important. 
We plan to investigate the properties of such model atmospheres in future studies.

\section*{Data availability}

Full set of spectral energy distributions of NS model atmospheres  for chemical compositions \texttt{ash1}--\texttt{ash4} as well as for pure iron considered in the paper are available at Zenodo  (\href{https://zenodo.org/records/20035991}{https://zenodo.org/records/20035991}). That record also presents  the best-fit parameters with the analytical model from Tables~\ref{tab:ash1}--\ref{tab:ash40}.

\begin{acknowledgements}
VFS was supported by the German Research Foundation (DFG) grant WE\,1312/59-1. 
JP acknowledges support from the  Research Council of Finland, the Centre of Excellence in Neutron-Star Physics (project 374064). 

\end{acknowledgements}


\bibliographystyle{aa}
\bibliography{ah}  

\begin{appendix} 
\section{Comparison with pure iron atmospheres} \label{app:iron}

In addition to the chemical compositions presented in this paper, {\sf ash1}–{\sf ash4}, we also computed model atmospheres composed of pure iron. Such atmospheres are clearly unlikely to occur in reality and therefore have primarily academic interest. Nevertheless, we present them for two reasons. First, to compare their properties with the previously published pure-iron models of \citet{Nattila.etal:15} and to demonstrate the changes introduced by the improved atmosphere calculations, in particular due to the inclusion of spectral lines and photoionization from excited levels. Second, these models illustrate the maximum optical depth that an absorption edge in X-ray burster spectra can reach. This can be properly assessed only in a chemically homogeneous atmosphere composed of a single heavy element near the iron peak, since a mixture of heavy elements (even the iron-dominated {\sf ash4} composition) tends to smooth the spectral features.

\begin{figure}[h]
\begin{center}  
\includegraphics[width=  0.78\columnwidth]{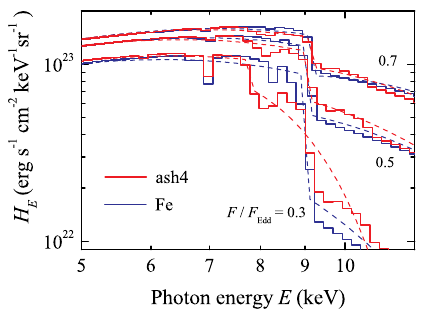}
\caption{\label{fig:a1} 
Comparison of spectra of model atmospheres consisting of pure iron (blue curves) and chemical composition {\sf ash4}  (red curves), calculated for the same relative fluxes and gravity ($\log g = 14.3$). The best spectral fits are also shown with dashed curves. }
\end{center} 
\end{figure}

\begin{figure} 
\begin{center}  
\includegraphics[width=  0.78\columnwidth]{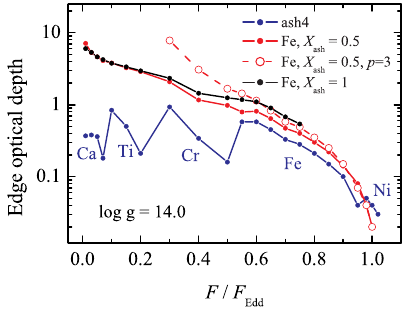}
\caption{\label{fig:a2} 
Evolution of the  optical depth at the threshold $\tau_{\rm th}$  with the relative flux for the models with iron dominated chemical composition {\sf ash4} and pure iron with $X_{\rm ash}= 0.5$ and 1. 
The effect of fixing the absorption exponent at $p=3$ is also shown. 
The fit parameters were obtained for $\log g = 14.0$. 
For the model {\sf ash4} the absorption edges of different chemical elements dominate, and they are marked.}
\end{center} 
\end{figure}

\begin{figure} 
\begin{center}  
\includegraphics[width=  0.78\columnwidth]{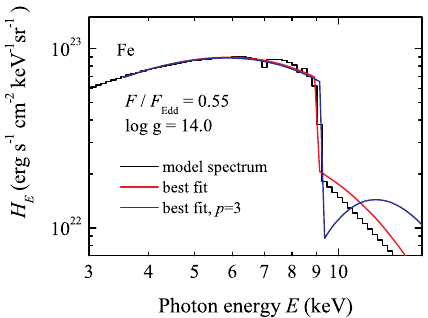}
\caption{\label{fig:a3} 
Best-fits of the atmosphere model spectrum with various approximations. 
The model spectrum  is shown with the black solid line. 
The best-fit function\,\eqref{eq:FwBE} with a free absorption exponent $p$ is shown with the red curve, while the fit with fixed $p=3$ is presented with the blue curve. }
\end{center} 
\end{figure}

The presence of significant amounts of nickel and chromium in the {\sf ash4} mixture leads to a blurring of the iron absorption edge due to the absorption edges of chromium and nickel (see Fig.\,\ref{fig:a1}). 
Moreover, with a decrease in the relative flux, and hence the effective temperature, the role of the most noticeable absorption edge accordingly passes from iron to elements with lower ionization energy, i.e. chromium, titanium, etc. (see Fig.\,\ref{fig:a2}). 
This proceeds in such a way that the optical depth of the main absorption edge remains relatively small and does not exceed unity, $\tau_{\rm th} <1$. 
On the contrary, in the case of pure iron atmospheres, the main absorption edge remains the same, and the optical thickness at the absorption threshold can significantly exceed unity (see again Fig.\,\ref{fig:a2}). 
We also note that the use of the \textsc{xspec} procedure \texttt{edge} (where the absorption exponent is fixed at $p=3$) in fitting the observed burster spectra leads to an overestimation of the optical depth at the photoionization threshold (see Fig.\,\ref{fig:a3}).

\begin{figure}[h]
\begin{center}  
\includegraphics[width=  0.78\columnwidth]{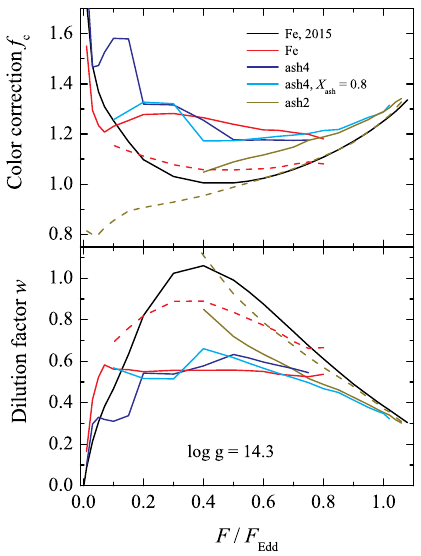}
\caption{\label{fig:a4} 
Evolution of the fit parameters with the relative flux. 
We considered here models with chemical composition {\sf ash4} ($X_{\rm ash} =1$ and 0.8), {\sf ash2}, and pure iron. 
The black solid curves shows the parameters for pure iron models from \citet{Nattila.etal:15}. 
The dashed curves correspond to the fits with diluted blackbody without an absorption edge. }
\end{center} 
\end{figure}

It is of interest to compare the main parameters of spectral fitting, color correction $f_{\rm c}$ and dilution factor $w$.
Such a comparison is presented in Fig.\,\ref{fig:a4}. 
We considered both models consisting of pure iron and models calculated for other mixtures.
The dependencies obtained by \citet{Nattila.etal:15} for pure iron atmospheres are also shown. 
Model spectra computed by \citet{Nattila.etal:15} were fitted by a diluted blackbody  without any absorption edge taken into account. 
Therefore, for models consisting of pure iron and a mixture of {\sf ash2}, we also presented (with dashed curves) the results of fitting with a simple diluted blackbody without taking into account the absorption edge.

We note that the color correction and dilution factors significantly differ if they were obtained using a diluted blackbody fit with and without an absorption edge.  
In particular, the color correction is higher and the dilution factor is lower if the absorption edge is included. 
It is interesting that the dependencies obtained for the {\sf ash2} mixture without taking into account the absorption edge coincide with those obtained by \citet{Nattila.etal:15} for pure iron for sufficiently high relative fluxes, $F/F_{\rm Edd} > 0.5$.

The color correction and dilution factors obtained for mixtures {\sf ash3}, {\sf ash4}, and pure iron depend weakly on the relative flux if it is high enough,  $F/F_{\rm Edd}>0.4$. 
In this work we found that pure iron atmospheres are unstable above a relative flux of about 0.75, although the previous work by \citet{Nattila.etal:15} calculated pure iron models up to much higher luminosities. 
The reason for this is that here we  took into account additional opacity, namely photoionization from excited levels and spectral lines.

\newpage 

\section{Fitting parameters} \label{app:fits}

Here we present the results of fitting the spectra of the calculated model atmospheres with different chemical compositions.
Two fitting functions were used: diluted blackbody and diluted blackbody with one absorption edge (see Eq.\,\eqref{eq:FwBE}). 
In the second case, the absorption edge threshold energy $E_{\rm th}$,  its optical thickness at the threshold $\tau_{\rm th}$, and the absorption exponent $p$, which determines the shape of the fit at photon energies exceeding $E_{\rm th}$,  are added to the diluted blackbody fitting parameters, color correction $f_{\rm c}$ and dilution factor $w$.

\begin{figure} 
\begin{center}  
\includegraphics[width=  0.78\columnwidth]{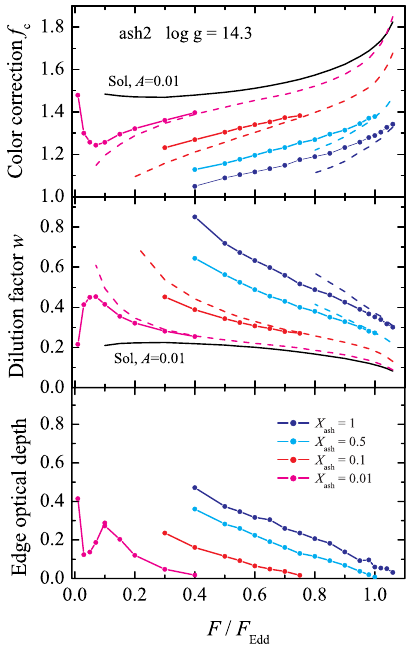}
\caption{\label{fig:B1} 
Evolution of the fit parameters with the relative bolometric flux for models with chemical composition {\sf ash2} and its mixing with a plasma of solar H/He mix and
sub-solar heavy elements abundances ($A=0.01$). The parameters of the fit  with a diluted blackbody without an absorption edge are shown with dashed lines.
For comparison, the evolution of the fit parameters for models with $X_{\rm ash}=0$ are also shown with the solid black line. 
}
\end{center} 
\end{figure}

\begin{figure} 
\begin{center}  
\includegraphics[width=  0.78\columnwidth]{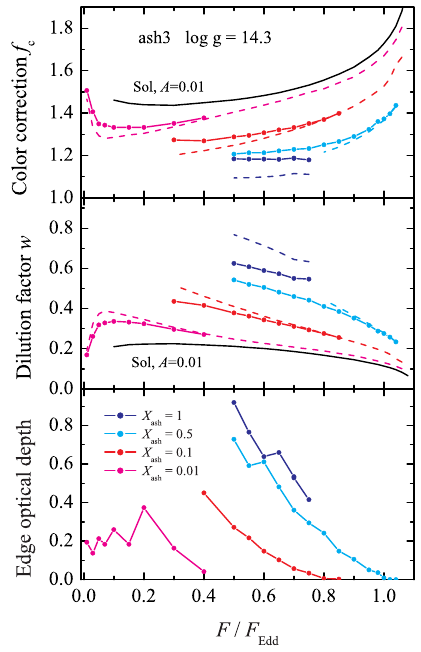}
\caption{\label{fig:B2} 
Same as Fig.\,\ref{fig:B1}, but for {\sf ash3}.
}
\end{center} 
\end{figure}

\begin{figure} 
\begin{center}  
\includegraphics[width=  0.78\columnwidth]{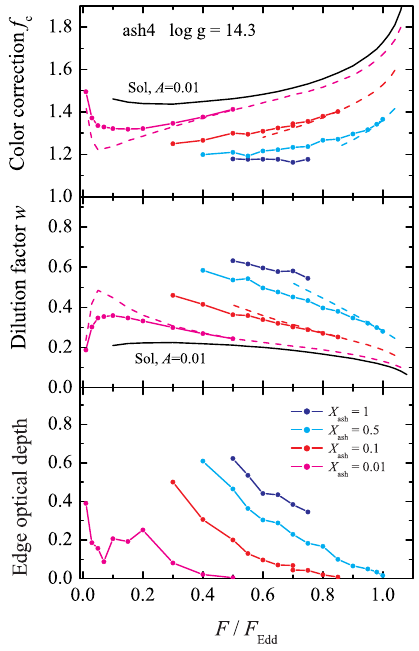}
\caption{\label{fig:B3} 
Same as Fig.\,\ref{fig:B1}, but for {\sf ash4}.
}
\end{center} 
\end{figure}

The results of fitting by both methods are presented in Figs.\,\ref{fig:B1}--\ref{fig:B3}. 
For the models with the relative ash abundances $X_{\rm ash} = 1, 0.5$, and  0.1, the results for high relative fluxes, where the absorption edges of the dominant chemical element (chromium for {\sf ash2}, nickel for {\sf ash3}, and iron for {\sf ash4}) are the main  absorption edges in the fits.  
If the absorption edge in the fit is negligible, only the results without the edge are shown (dashed curves).

The results of fitting the spectra of model atmospheres consisting exclusively of thermonuclear ashes are also presented in the tables.
Model spectra of atmospheres consisting of helium-dominated ashes (mix {\sf ash1}) were fitted only with a diluted blackbody (see Table\,\ref{tab:ash1}).
For the remaining mixtures of thermonuclear ashes, the results of fitting with a diluted blackbody are given, taking into account the absorption edge (Tables\,\ref{tab:ash2}--\ref{tab:ash4}). 
To illustrate the potential importance of the gravity value, models with $\log g = 14.0$ were calculated too. 
Their emergent spectra were fitted by a diluted blackbody taking into account the absorption edge (see Table\,\ref{tab:ash40}). 
Tables with fitting results for models consisting of mixtures of thermonuclear ashes and plasma of solar chemical composition ($X_{\rm ash} = 0.5, 0.1$, and 0.01) can be provided upon request.

 \begin{table}[h]
\caption{Fitting parameters of the models with the chemical composition {\sf ash1}, $X_{\rm ash}=1$, and $\log g = 14.3$.  
 \label{tab:ash1} 
 }
{\footnotesize
\begin{center} 
\begin{tabular}{ccc}
 \hline\hline
  $F/F_{\rm Edd}$  & $w$ & $f_{\rm c}$ \\ 
\hline

   1.080 &   0.099 &   1.802    \\
   1.060 &   0.126 &   1.685    \\
   1.040 &   0.151 &   1.604    \\
   1.020 &   0.163 &   1.570    \\
   1.000 &   0.175 &   1.536    \\
   0.980 &   0.185 &   1.514    \\
   0.950 &   0.200 &   1.484    \\
   0.900 &   0.221 &   1.444    \\
   0.850 &   0.234 &   1.422    \\
   0.800 &   0.246 &   1.403    \\
   0.750 &   0.258 &   1.387    \\
   0.700 &   0.264 &   1.379    \\
   0.650 &   0.269 &   1.373    \\
   0.600 &   0.275 &   1.364    \\
   0.550 &   0.277 &   1.362    \\
   0.500 &   0.277 &   1.362    \\
   0.400 &   0.274 &   1.366    \\
   0.300 &   0.266 &   1.377    \\
   0.200 &   0.252 &   1.396    \\
   0.150 &   0.239 &   1.412    \\
   0.100 &   0.224 &   1.435    \\
   0.070 &   0.208 &   1.460    \\
   0.050 &   0.193 &   1.485    \\
   0.030 &   0.170 &   1.526    \\
   0.010 &   0.109 &   1.675    \\
\hline
\end{tabular}
\end{center}
}
\end{table}

 \begin{table}
\caption{Fitting parameters of the models with the chemical composition {\sf ash2}, $X_{\rm ash}=1$, and $\log g = 14.3$.  
 \label{tab:ash2} 
 }
{\footnotesize
\begin{center} 
\begin{tabular}{cccccr}
 \hline\hline
  $F/F_{\rm Edd}$  & $w$ & $f_{\rm c}$ & $E_{\rm th}$   &  $\tau_{\rm th}$ & $p$\\ 
\hline

   1.060 &   0.302 &   1.342 &   7.89 &   0.03 &   0.94  \\
   1.040 &   0.318 &   1.327 &   7.97 &   0.05 &   0.94  \\
   1.020 &   0.340 &   1.305 &   7.83 &   0.05 &   0.33  \\
   1.000 &   0.353 &   1.289 &   7.85 &   0.06 &   0.41  \\
   0.980 &   0.367 &   1.279 &   7.93 &   0.10 &   0.83  \\
   0.950 &   0.390 &   1.258 &   7.91 &   0.09 &   0.54  \\
   0.900 &   0.426 &   1.232 &   7.81 &   0.14 &   0.61  \\
   0.850 &   0.462 &   1.205 &   7.86 &   0.18 &   1.20  \\
   0.800 &   0.488 &   1.189 &   7.78 &   0.21 &   1.14  \\
   0.750 &   0.517 &   1.175 &   7.67 &   0.24 &   0.98  \\
   0.700 &   0.560 &   1.148 &   7.84 &   0.26 &   1.18  \\
   0.650 &   0.598 &   1.132 &   7.74 &   0.30 &   1.23  \\
   0.600 &   0.633 &   1.117 &   7.91 &   0.32 &   1.10  \\
   0.550 &   0.673 &   1.104 &   7.72 &   0.35 &   1.20  \\
   0.500 &   0.719 &   1.089 &   7.57 &   0.37 &   0.75  \\
   0.400 &   0.850 &   1.049 &   7.74 &   0.47 &   1.04  \\
   0.300 &   0.751 &   1.137 &   6.48 &   0.25 &  $-$2.88  \\
   0.200 &   0.892 &   1.106 &   6.56 &   0.36 &  $-$3.49  \\
   0.150 &   0.547 &   1.364 &   5.28 &   0.16 &  $-$5.50  \\
   0.100 &   0.505 &   1.417 &   5.31 &   0.30 &  $-$5.22  \\
   0.070 &   0.480 &   1.441 &   5.24 &   0.46 &  $-$4.78  \\
   0.050 &   0.673 &   1.275 &   5.27 &   0.58 &  $-$5.90  \\
   0.030 &   0.157 &   2.093 &   4.71 &   0.29 & $-$10.78  \\
   0.010 &   0.036 &   3.035 &   4.55 &   0.53 & $-$14.28  \\
\hline
\end{tabular}
\end{center}
}
\end{table}

\begin{table}
\caption{Fitting parameters of the models with the chemical composition {\sf ash3}, $X_{\rm ash}=1$, and $\log g = 14.3$.  
 \label{tab:ash3} 
 }
{\footnotesize
\begin{center} 
\begin{tabular}{cccccr}
 \hline\hline
  $F/F_{\rm Edd}$  & $w$ & $f_{\rm c}$ & $E_{\rm th}$   &  $\tau_{\rm th}$ & $p$\\ 
\hline

   0.750 &   0.547 &   1.179 &  10.62 &   0.41 &   0.27  \\
   0.700 &   0.551 &   1.187 &  10.64 &   0.53 &   0.74  \\
   0.650 &   0.574 &   1.182 &  10.77 &   0.66 &   0.70  \\
   0.600 &   0.590 &   1.183 &  10.54 &   0.64 &  $-$0.40  \\
   0.550 &   0.608 &   1.183 &  10.44 &   0.77 &  $-$0.49  \\
   0.500 &   0.626 &   1.184 &  10.61 &   0.93 &  $-$0.61  \\
   0.400 &   0.573 &   1.247 &   9.28 &   0.40 &  $-$5.63  \\
   0.300 &   0.559 &   1.269 &   9.00 &   0.51 &  $-$6.52  \\
   0.200 &   0.410 &   1.423 &   7.60 &   0.34 &  $-$5.80  \\
   0.200 &   0.425 &   1.403 &   7.83 &   0.54 &  $-$3.93  \\
   0.150 &   0.401 &   1.419 &   7.58 &   0.67 &  $-$3.96  \\
   0.100 &   0.435 &   1.358 &   7.47 &   1.19 &  $-$2.92  \\
   0.070 &   0.318 &   1.494 &   6.60 &   0.46 &  $-$6.13  \\
   0.050 &   0.340 &   1.428 &   6.57 &   0.51 &  $-$6.49  \\
   0.030 &   0.290 &   1.444 &   6.42 &   0.62 &  $-$5.63  \\
   0.010 &   0.096 &   1.782 &   5.52 &   0.38 &  $-$3.63  \\
\hline
\end{tabular}
\end{center}
}
\end{table}

 \begin{table}
\caption{Fitting parameters of the models with the chemical composition {\sf ash4}, $X_{\rm ash}=1$, and $\log g = 14.3$.  
 \label{tab:ash4} 
 }
{\footnotesize
\begin{center} 
\begin{tabular}{cccccr}
 \hline\hline
  $F/F_{\rm Edd}$  & $w$ & $f_{\rm c}$ & $E_{\rm th}$   &  $\tau_{\rm th}$ & $p$\\ 
\hline

   0.750 &   0.545 &   1.176 &   9.11 &   0.35 &   0.12 \\
   0.700 &   0.582 &   1.162 &   9.27 &   0.38 &  $-$0.09  \\
   0.650 &   0.579 &   1.176 &   9.11 &   0.43 &   0.08 \\
   0.600 &   0.596 &   1.177 &   9.08 &   0.44 &  $-$0.63  \\
   0.550 &   0.616 &   1.176 &   9.12 &   0.54 &  $-$0.61  \\
   0.500 &   0.632 &   1.178 &   9.02 &   0.62 &  $-$0.72  \\
   0.400 &   0.577 &   1.255 &   7.72 &   0.20 &  $-$5.63  \\
   0.300 &   0.538 &   1.316 &   7.80 &   0.38 &  $-$5.63  \\
   0.200 &   0.543 &   1.319 &   7.67 &   0.86 &  $-$3.59  \\
   0.150 &   0.337 &   1.579 &   6.60 &   0.18 &  $-$9.33  \\
   0.100 &   0.310 &   1.581 &   6.51 &   0.35 &  $-$7.35  \\
   0.070 &   0.319 &   1.526 &   6.60 &   0.92 &  $-$3.65  \\
   0.050 &   0.330 &   1.473 &   6.47 &   1.03 &  $-$3.84  \\
   0.030 &   0.297 &   1.467 &   6.51 &   1.25 &  $-$4.49  \\
   0.010 &   0.096 &   1.854 &   5.55 &   0.36 &  $-$5.91  \\
\hline
\end{tabular}
\end{center}
}
\end{table}

\begin{table}
\caption{Fitting parameters of the models with the chemical composition {\sf ash4}, $X_{\rm ash}=1$, and $\log g = 14.0$.  
\label{tab:ash40} }
{\footnotesize
\begin{center} 
\begin{tabular}{cccccc}
 \hline\hline
  $F/F_{\rm Edd}$  & $w$ & $f_{\rm c}$ & $E_{\rm th}$   &  $\tau_{\rm th}$ & $p$\\ 
\hline
   0.750 &   0.495 &   1.226 &   9.18 &   0.29 &  $-$1.02 \\
   0.700 &   0.531 &   1.212 &   9.24 &   0.43 &  $-$0.83 \\
   0.650 &   0.586 &   1.184 &   9.26 &   0.65 &  $-$0.64 \\
   0.600 &   0.590 &   1.195 &   9.11 &   0.60 &  $-$1.71 \\
   0.550 &   0.620 &   1.187 &   9.06 &   0.71 &  $-$1.86 \\
   0.500 &   0.550 &   1.257 &   7.88 &   0.32 &  $-$4.25 \\
   0.400 &   0.532 &   1.292 &   7.78 &   0.51 &  $-$4.25 \\
   0.300 &   0.507 &   1.317 &   7.68 &   0.95 &  $-$3.56 \\
   0.200 &   0.328 &   1.532 &   6.47 &   0.44 &  $-$4.25 \\
   0.150 &   0.320 &   1.516 &   6.49 &   0.50 &  $-$5.66 \\
   0.100 &   0.325 &   1.468 &   6.47 &   0.95 &  $-$3.76 \\
   0.070 &   0.190 &   1.733 &   5.55 &   0.20 &  $-$8.75 \\
   0.050 &   0.190 &   1.675 &   5.63 &   0.38 &  $-$6.25 \\
   0.030 &   0.139 &   1.743 &   5.54 &   0.45 &  $-$5.29 \\
   0.010 &   0.054 &   1.975 &   5.59 &   0.52 &  $-$4.41 \\
\hline
\end{tabular}
\end{center}
}
\end{table}

\end{appendix}

\end{document}